\documentclass[11pt]{article}
\usepackage[utf8]{inputenc} 
\usepackage[T1]{fontenc}
\usepackage{amsmath, amssymb, bm}  
\usepackage{graphicx}        
\usepackage{siunitx}         
\usepackage{xcolor}          
\usepackage{authblk}         
\usepackage{geometry}         
\title{Evaluation of the Transfer Matrix of a Plasma Ramp with Squared Cosine Shape via an Approximate Solution of the Mathieu Differential Equation}

\author[1,*]{S. Romeo}
\author[1]{A. Biagioni}
\author[1,2]{L. Crincoli}
\author[1]{A. Del Dotto}
\author[1]{M. Ferrario}
\author[1]{A. Giribono}
\author[1,3]{G. Parise}
\author[4]{A.R. Rossi}
\author[2,5]{G.J. Silvi}
\author[1]{C. Vaccarezza}

\affil[1]{Laboratori Nazionali di Frascati, Via Enrico Fermi 54, 00044 Frascati (Rome), Italy}
\affil[2]{Sapienza University, Piazzale Aldo Moro 5, 00185 Rome, Italy}
\affil[3]{Department of Physics, University of Rome "Tor Vergata", Via della Ricerca Scientifica 1, 00133 Rome, Italy}
\affil[4]{INFN-Milan, Via Celoria 16, 20133 Milan, Italy}
\affil[5]{Sezione di Roma - INFN, Piazzale Aldo Moro 5, 00185 Rome, Italy}
\affil[*]{\texttt{stefano.romeo@lnf.infn.it}}

\date{} 

\begin{document}

\maketitle

\begin{abstract}
The high longitudinal electric fields generated in plasma wakefields are very attractive for a new generation of high gradient plasma based accelerators. On the other hand, the strong transverse fields increase the demand for a proper matching device in order to avoid the spoiling of beam transverse quality. A solution can be provided by the use of a plasma ramp, a region at the plasma injection/extraction with smoothly increasing/decreasing plasma density. The transport of a beam inside a plasma ramp, beside its parameters, depends on the profile of the ramp itself. Establishing the transfer matrix for a plasma ramp represents a very useful tool in order to evaluate the beam evolution in the plasma. In this paper a study of a cosine squared ramp is presented. An approximate solution of the transverse equation of motion is evaluated and exploited to provide a simple transfer matrix for the plasma ramp. The transfer matrix is then employed to demonstrate that this kind of ramp has the effect to minimize the emittance growth due to betatron dephasing. The behavior of a squared cosine plasma ramp will be compared with an experimentally measured plasma ramp profile in order to validate the applicability of the transfer matrix to real cases.
\end{abstract}

\section{Introduction}

Plasma-based devices have been proven to be a valid candidate for the development of a new generation of accelerating machines~\cite{hogan2010plasma} due to the high accelerating gradients, up to tens of GV/m~\cite{leemans2006gev,blumenfeld2007energy,litos2014high}, that they are able to produce. The main actual task of plasma wakefield acceleration is the preservation of the beam quality during the acceleration process both in terms of emittance and energy spread. Recent results have proven that modern plasma modules and schemes are able to accelerate bunches that can be fully characterized~\cite{shpakov2021first,pompili2021energy} and exploited in order to pilot a free electron laser~\cite{wang2021free,pompili2022free}, a device that notoriously requires high quality bunches only.\\
Due to these reasons, emittance preservation is a main topic for the design of plasma based facilities~\cite{assmann2020eupraxia} and several efforts were performed in order to evaluate optimal conditions for beam transport inside plasma \cite{barov1994propagation}. Unfortunately, in most cases of interest, the transverse beam matching inside a plasma requires a focusing at the injection from few micrometers up to sub-micrometer scale in order to completely avoid emittance growth. A plasma ramp \cite{chen1990plasma} is usually referred to as a section of plasma where the density varies smoothly. This is a partial definition that ignores the studies performed on low density constant plasma ramps~\cite{ariniello2019transverse}. In this case, the increase of the plasma density is totally sharp, but, for all purposes, this can be considered a ramp. A more correct definition is that a plasma ramp is a section of the plasma profile that introduces a smooth variation of transverse beam parameters. It has been widely demonstrated that for several ramp shapes \cite{floettmann2014adiabatic,xu2016physics,tomassini2015matching,ariniello2019transverse} the presence of a plasma ramp is very helpful to reduce emittance growth inside plasma. However, these studies all refer to adiabatic ramps \cite{chen1990plasma,floettmann2014adiabatic,ariniello2019transverse}, a kind of ramp where the focusing strength varies slowly compared to the oscillation wavelength of the bunch. This feature can be mathematically expressed as \cite{ariniello2019transverse}
\begin{equation}
    \mathcal{A}=\frac{1}{4\eta^{3/2}(z)}\left\lvert\frac{d\eta(z)}{dz}\right\lvert
    \label{eqn:adiab}\ll1,
\end{equation}
where $\mathcal{A}$ is defined as adiabaticity parameter and $\eta(z)$ is the plasma focusing strength evolving along the direction of the bunch $z$. On the other side, analytical and numerical solutions have been derived for a limited number of non-adiabatic ramps with different shapes \cite{xu2016physics,ariniello2019transverse}. Adiabatic ramps are the most promising in terms of performances, but less appealing in terms of applications. Tapered profiles have been discussed in order to experimentally obtain the smoothly varying shape required for adiabatic focusing \cite{filippi2018tapering,rosenzweig2018adiabatic}. From a practical point of view, the introduction of a long adiabatic ramp at the injection and extraction of a plasma channel severely reduces the average accelerating field of the whole plasma section, frustrating the efforts for increasing the accelerating gradient that is the strong point of plasma acceleration itself. Besides, in non confined plasma, as for example from capillary discharge, non-adiabatic ramps are naturally formed \cite{filippi2016spectroscopic} without further technological effort. As pointed out from Ariniello et al. \cite{ariniello2019transverse}, in most cases the longitudinal shapes of the plasma ramps with an analytical solution are described by discontinuous functions or present a discontinuity of the first longitudinal derivative at the junction point between the ramp and the flat-top channel. This kind of ramp is clearly unphysical, despite most of the derived solutions can still be considered as good approximations to reality. The work proposed in this paper is focused on the study of ramps with squared cosine shape, a class of plasma ramps that never presents a fully adiabatic behavior for realistic ramp lengths. The evaluation of the transfer matrix for this ramp, that is a key byproduct of the analytical integration of the motion equations, requires the solution of Hill's differential equation when the focusing term is a sinusoidal function, an equation also known as Mathieu differential equation \cite{abramowitz1964handbook}. An approximated solution will be provided for this equation in the case of medium range non-adiabatic ramp. The solutions will be applied in order to prove the stabilizing effect of the ramp in terms of transverse emittance growth. Finally, the transfer matrix will be applied to find the matching of a measured plasma profile in order to prove the possibility to practically obtain a plasma ramp with the characteristics and the behavior of a squared cosine ramp.
\section{Transverse matching condition of a plasma channel at the plateau}
The transverse matching condition for a bunch injected into a focusing channel can be mathematically expressed in several ways, which are equivalent to each other due to the relationship that is intrinsic to the envelope equation \cite{wiedemann2015particle}
\begin{equation}
    \beta''(z)+2k_{ext}^2(z)\beta(z)=\frac{2}{\beta(z)}+\frac{\beta'^2(z)}{2\beta(z)},
    \label{eqn:env1}
\end{equation}
where $k_{ext}^2(z)$ is the normalized focusing strength and $\beta$ is the Twiss $\beta$-function. The envelope equation is derived by means of the Courant-Snyder equation in combination with the equations of motion and is valid assuming the paraxial approximation, a low energy spread and constant emittance of the bunch. The $\beta$-function expressed in Eq.(\ref{eqn:env1}) is derived from the equations of motions and describes the ability of a transport line to focus a particle bunch, given the required boundary conditions. We neglected the subscript $x,y$ since this equation is valid for an arbitrary choice of transverse directions and in our work we will assume cylindrical symmetry. If the focusing force is constant along the channel, the matching condition is a trivial quasi-stationary solution of Eq.(\ref{eqn:env1})
\begin{equation}
    \begin{split}
        \beta(0)&=\frac{1}{k_{ext}}\\
        \beta'(0)&=0\\
        \beta''(0)&=0.
    \end{split}
\end{equation}
It is easily verified that, assuming any two of these equations, the third one follows immediately from Eq.(\ref{eqn:env1}). The focusing term in our case of interest is given by the varying plasma focusing force. Its description is not a trivial task since plasma wave is formed as a result of the interaction of the driving pulse, either a driving bunch or a laser pulse, with the background bulk. The effects of the plasma ionization, the formation of the plasma wave and the ion motion will be neglected. From now on, it will be assumed that the bunch transported inside plasma is totally contained in a plasma bubble in non-linear regime. The focusing force, under these assumptions, can be represented by the ion column model \cite{barov1994propagation}. The plasma density $n_p(z,r)$ is generally not constant in either longitudinal or transverse direction, but for the purposes of this work, the transverse dependency of the plasma density will be neglected as well. The normalized focusing strength of the ion channel is
\begin{equation}
    k^2_{ext}(z)=\frac{k_p^2(z)}{2\gamma},
    \label{eqn:plasma1}
\end{equation}
where $\gamma$ is the bunch average Lorentz factor and $k_p=[e^2n_p(z)/\epsilon_0 m_e c^2]^{1/2}$ is the plasma wave number. The matching conditions for a plasma plateau with a constant density is then
\begin{equation}
    \beta_0=\frac{\sqrt{2\gamma}}{k_p},
    \label{eqn:betam}
\end{equation}
that for typical plasma and beam parameters is of the order of 1 $\mu m$ or smaller. An input plasma ramp is a device where the density $n_p(z)$ is not constant, but rises from $0$ up to the nominal value $n_0$ of the channel. In order to simplify any further evaluation, it will be assumed that the energy variation of the bunch inside the ramp itself is negligible. In order to verify this assumption, the longitudinal field inside the ramp must be estimated. The average accelerating gradient acting on a ramp can be evaluated from the shape of the ramp, having an equation that links the accelerating gradient to the plasma density. The most natural connection would be given by the wavebreaking limit~\cite{rosenzweig1988trapping} $E_z[$\SI{}{\electronvolt}$]=96(n_p[$\SI{}{\per\cubic\centi\meter}$])^{1/2}$, that describes the maximum accelerating gradient that can be generated in a plasma wave without destroying its oscillating behavior. In a linear ramp arising from $0$ to $n_0$ in a length $L$, the average field would be given by $2/3$ of the wavebreaking limit evaluated at the end of the ramp. Thus, the assumption of negligible acceleration is certainly valid as long as $\gamma_0\gg64L[$\SI{}{\meter}$](n_0[$\SI{}{\per\cubic\centi\meter}$])^{1/2}/m_e[$\SI{}{\electronvolt}$]$ where $m_e$ is the electron mass expressed in electronvolt. However, the usage of the wavebreaking limit is an extreme feature, since the plasma accelerators rarely reach fields that are even comparable. It is safe to state that the average accelerating gradient is usually at least one order of magnitude lower. The overall complexity of this aspect suggests that it is best verified with numerical simulations, afterwards. From now on, the work takes into account bunches with an energy of several hundreds of MeVs, a plasma density around $10^{16}$\SI{}{\per\cubic\centi\meter} and ramps with a length of few centimeters, that have been tested to be well below this limit. Anyway, the assumption allows from Eq.(\ref{eqn:plasma1}) to state that $n_p$ and $k_{ext}^2$ are equal, except for a constant, so the focusing strength grows from $0$ up to a nominal value as well.  We can write $k_{ext}^2(z)=\eta(z)$ where we assume that $\eta(z)$ is a continuous and differentiable function in an arbitrary domain since the behavior of realistic plasma ramps require these assumptions. Further, from Eq.(\ref{eqn:betam}), one can also write $\eta(0)=1/\beta_0^2$. With a proper configuration, a plasma ramp can help to relax the matching conditions at the entrance of the plasma. We will assume that the beam is traveling with a negative momentum, namely that beam starts from $z_0>0$ at the beginning of the ramp, traveling up to $z=0$ where the ramp ends and start traveling inside the channel located at $z<0$ for an undefined length. The reverse transfer matrix of the channel, thus the behavior of a beam moving from $z=0$ from left to right, will be computed. Based on our previous considerations, a matched beam at $z=0$ will have the following Twiss vector
\begin{equation}
    \begin{bmatrix}
    \beta(0)\\
    \alpha(0)\\
    \gamma(0)\\
    \end{bmatrix}
    =
    \begin{bmatrix}
    \beta_0\\
    0\\
    1/\beta_0\\
    \end{bmatrix};
    \label{eqn:cond}
\end{equation}
If $C(z)$ and $S(z)$ are the even and odd solutions to Hill's equation inside the plasma ramp respectively
\begin{equation}
   y''(z)+\eta(z)y(z)=0;
   \label{eqn:hill}
\end{equation}
the transfer matrix for the Twiss parameters~\cite{wiedemann2015particle} is 
\begin{equation}
    \begin{bmatrix}
    \beta(z)\\
    \alpha(z)\\
    \gamma(z)\\
    \end{bmatrix}
    =
    \begin{bmatrix}
    C^2 & -2CS & S^2\\
    -CC' & CS'+C'S & -SS'\\
    C'^2 & -2C'S' & S'^2\\
    \end{bmatrix}
    \begin{bmatrix}
    \beta_0\\
    0\\
    1/\beta_0\\
    \end{bmatrix},
    \label{eqn:bmatrix}
\end{equation}
where, for sake of clarity, we omitted the dependency of the functions on $z$. So we have a new set of equations for the Twiss functions at any point in the ramp
\begin{equation}
    \begin{split}
        \beta(z)&=C^2 \beta_0+S^2/\beta_0\\
        \alpha(z)&=-CC' \beta_0 -SS'/\beta_0\\
        \gamma(z)&=C'^2 \beta_0+S'^2/\beta_0.\\
    \label{eqn:twiss}
    \end{split}
\end{equation}
In the assumption of a continuous and differentiable $\eta(z)$, the solutions to Hill's equation are continuous, differentiable twice with continuous second derivative.
\section{Symmetric ramps}
The most simple case occurs when the ramp function $\eta(z)$ is an even function. A linear differential equation of $n^{th}$ order
\begin{equation}
    \sum\limits_{k=0}^{n} u_k(z)\frac{d^ky}{dz^k}=t(z);
\end{equation}
is symmetric as long as the even terms, $u_{k=2j}(z)$ and $t(z)$, and the odd terms, $u_{k=2j+1}(z)$, have opposite parities. Recalling Eq.(\ref{eqn:hill}) one recognizes that $t(z)=0$ and $u_2(z)=1$ are both even functions while $u_1=0$ is an odd function. If we assume that $u_0=\eta(z)$ is an even function too, Eq.(\ref{eqn:hill}) is a symmetric differential equation of the second order that admits an even, $C(z)$, and an odd, $S(z)$, solution. These functions can be normalized so that $C(0)=1$ and $S'(0)=1$. The boundary conditions at $z=0$ can be summarized as follows
\begin{align} \label{eqn:Bcond}
        \beta(0)&=\beta_0 &\alpha(0)&=0 \nonumber\\
        \gamma(0)&=1/\beta_0 &\beta''(0)&=0.
\end{align}
  We recall that the derivative of an even function is odd and vice versa. Since $C$ and $S$ are assumed to be differentiable, we can also write the following boundaries
\begin{align} \label{eqn:CScond}
        C(0)&=1 &C'(0)&=0 \nonumber\\
        S(0)&=0 &S'(0)&=1.
\end{align}
It is easy to verify that substituting Eq.(\ref{eqn:CScond}) into Eq.(\ref{eqn:twiss}), the conditions Eq.(\ref{eqn:Bcond}) are automatically respected, thus for an even function the only requirement for boundary is re-normalization of the solutions.
\section{Squared cosine plasma ramp}
\begin{figure}[h!]
    \centering
    \includegraphics[width=0.5\linewidth]{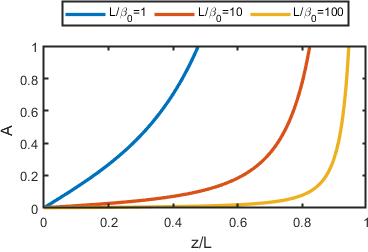}
    \caption{Evolution of the adiabatic parameter as a function of $z/L$ for different values of $L/\beta_0$.}
    \label{fig:adiab}
\end{figure}

The symmetric choice of the form for $\eta(z)$ is the following
\begin{equation}
\begin{cases}
    \eta(z)=\frac{1}{\beta_0^2}\hspace{5mm} &$for$\ z\leq0;\\
      \eta(z)=\frac{1}{\beta_0^2} \cos^2{\left(\frac{\pi z}{2L}\right)}\hspace{5mm} &$for$\ 0\leq z \leq L;\\
    \eta(z)=0 &$for$\ z>L;
\end{cases}
\end{equation}
that is a continuous, differentiable function with continuous first derivative, for which it is also possible to find a set of Twiss parameters at injection that guarantee the matching at $z=0$. For this kind of ramp, the adiabaticity parameter from Eq.(\ref{eqn:adiab}) can be evaluated as
\begin{equation}
    \mathcal{A}=\frac{\pi \beta_0}{4 L}\frac{\sin\left(\frac{\pi z}{2L}\right)}{\cos^2\left(\frac{\pi z}{2L}\right)}.
    \label{eqn:adiab2}
\end{equation}
In Fig.(\ref{fig:adiab}) the value of $\mathcal{A}$ is shown as a function of $z/L$ from $0$ (plateau) to $1$ (end of the ramp) for different values of $L/\beta_0$. The adiabaticity parameter presents a divergence at $z=L$ and goes to $0$ at $z=0$, meaning that for any value of $L$ this kind of ramp is non-adiabatic. For high values of $L/\beta_0$, the values of $z/L$ where $\mathcal{A}\ll1$ increase, meaning that non-adiabatic behavior is concentrated in the first part of the ramp, getting adiabatic when the bunch approaches to the plateau. As a consequence, adiabatic approach is not possible and $C$ and $S$ shall be evaluated from Hill's equation.
The Hill's equation for the ramp can be written as
\begin{equation}
    \frac{d^2 y(z)}{d z^2}+\frac{1}{\beta_0^2}\cos^2\left(\frac{\pi z}{2L}\right) y(z)=0
\end{equation}
that, by the use of trigonometric identities and substitution of variables, can be turned into the following equation
\begin{equation}
    \frac{d^2 Y(v)}{d v^2}+ \frac{2L^2}{\pi^2\beta_0^2}\left[1+\cos(2v)\right]Y(v)=0,
    \label{eqn:ourmathi}
\end{equation}
where $v=\pi z/2L$. Eq.(\ref{eqn:ourmathi}) can be identified as a Mathieu differential equation (MDE)\cite{brimacombe2021computation}, whose general form is
\begin{equation}
    V''(v)+[a-2q\cos(2v)]V(v)=0.
    \label{mathimathi}
\end{equation}
Assuming periodic boundary conditions, analytical solutions of MDE exist only for given couples of $a$ and $q$. This does not occur in the description of a plasma ramp since the boundaries are given in Eq.(\ref{eqn:CScond}). The general solutions to MDE are known as Mathieu even $\mathcal{C}(a,q,v)$ and odd $\mathcal{S}(a,q,v)$ functions. As stated, the Mathieu functions can be normalized in order to respect the boundary conditions, so we can write the general solutions for the Hill's equations in the ramp as
\begin{equation}
\begin{cases}
    C(z)&=\frac{\mathcal{C}\left(\frac{2L^2}{\pi^2 \beta_0^2},-\frac{L^2}{\pi^2 \beta_0^2},\frac{\pi z}{2 L}\right)}{\mathcal{C}\left(\frac{2L^2}{\pi^2. \beta_0^2},-\frac{L^2}{\pi^2 \beta_0^2},0\right)};\\
    \\
    S(z)&=\frac{2 L}{\pi}\frac{\mathcal{S}\left(\frac{2L^2}{\pi^2 \beta_0^2},-\frac{L^2}{\pi^2 \beta_0^2},\frac{\pi z}{2 L}\right)}{\mathcal{S}'\left(\frac{2L^2}{\pi^2 \beta_0^2},-\frac{L^2}{\pi^2 \beta_0^2},0\right)};
\end{cases}
\label{eqn:nmathi}
\end{equation}
where $\mathcal{S}'$ is the Mathieu odd function derivative. The solution Eq.(\ref{eqn:nmathi}) respects all the criteria from the boundary conditions. The normalized solution can be expressed in the compact form
\begin{equation}
\begin{cases}
    c(\xi;R)&=C(z)=\frac{\mathcal{C}\left(2R^2,-R^2,\frac{\pi}{2}\xi\right)}{\mathcal{C}\left(2R^2,-R^2,0\right)};\\
    \\
    s(\xi;R)&=\frac{S(z)}{\beta_0}=2R\frac{\mathcal{S}\left(2R^2,-R^2,\frac{\pi}{2 }\xi\right)}{\mathcal{S}'\left(2R^2,-R^2,0\right)};\\
    \\
    c'(\xi;R)&=\beta_0 C'(z)= \frac{1}{2R} \frac{\mathcal{C}'\left(2R^2,-R^2,\frac{\pi}{2 }\xi\right)}{\mathcal{C}\left(2R^2,-R^2,0\right)};\\
    \\
    s'(\xi;R)&=S'(z)=\frac{\mathcal{S}'\left(2R^2,-R^2,\frac{\pi}{2}\xi\right)}{\mathcal{S}'\left(2R^2,-R^2,0\right)};
\end{cases}
\label{eqn:comp_mathi}
\end{equation}
where $\xi=z/L$ and $R=L/\pi\beta_0$ is the ratio between the length of the ramp and the betatron wavelength at the plateau. By inserting Eq.(\ref{eqn:comp_mathi}) in Eq.(\ref{eqn:bmatrix}) one can evaluate the evolution of Twiss parameters inside the ramp as
\begin{equation}
    \begin{split}
        \frac{\beta}{\beta_0}&=c^2(\xi)+s^2(\xi)\\
        \alpha&=-c(\xi)c'(\xi)-s(\xi)s(\xi)'.
    \end{split}
    \label{eqn:twiss_ev}
\end{equation}
It is important to notice Eq.(\ref{eqn:comp_mathi},\ref{eqn:twiss_ev}) only depend on parameter $R$, describing an entire class of ramps. The matching behavior of the ramps is exactly the same as long the $R$ parameter is the same.
\begin{figure}[h!]
    \centering
    \includegraphics[width=0.5\linewidth]{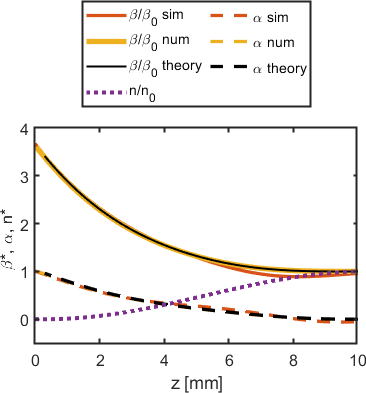}
    \caption{Comparison of the evolution of Twiss $\beta$-function and $\alpha$-function inside a plasma ramp (purple) evaluated by means of numerical simulation (red), numerical integration of envelope equation (yellow) and transfer matrix of the plasma ramp (black).}
    \label{fig:twiss_ev}
\end{figure}
In Fig. \ref{fig:twiss_ev} the evolution of the Twiss $\beta$-function and $\alpha$-function is shown, evaluated by use of Eq.s \eqref{eqn:comp_mathi} and \eqref{eqn:twiss_ev} compared to the numerical integration of envelope equation and Eq.(\ref{eqn:twiss_ev}) and a numerical simulation performed with the hybrid kinetic-fluid code Architect \cite{marocchino2016efficient}. The considered bunch energy is $E_0=500$ MeV, the peak plasma density is $n_0=10^{16}$cm$^{-3}$ and the ramp length is $10$mm. As can be clearly seen, Eq.(\ref{eqn:twiss_ev}) is substantially equivalent to numerical integration of envelope equation and they are both in good agreement with simulation results.

\section{Approximated transfer matrix}
\begin{figure}[h!]
    \centering
    \centering
    \includegraphics[width=0.5\linewidth]{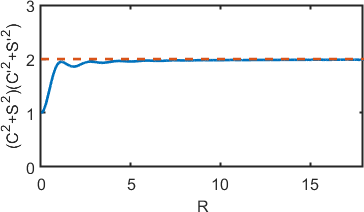}
    \caption{$(c^2+s^2)(c'^2+s'^2)$ as a function of $R$. The function saturates quickly at $2$ for values of $R>5$.}
    \label{fig:mathi3}
\end{figure}

\begin{figure}[h!]
    \centering
    \includegraphics[width=0.5\linewidth]{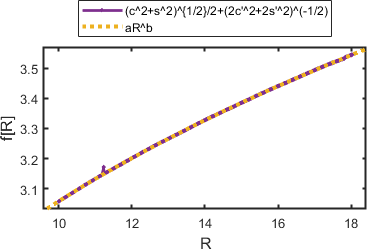}
    \caption{Comparison of the function $\sqrt{c^2+s^2}/2+\sqrt{2/({c'}^2+{s'}^2)}/2$ with the resulting fit $\sqrt{3} R^{1/4}$.}
    \label{fig:mathi4}
\end{figure}

By setting $\xi=1$ it is possible to describe the evolution of the bunch from the beginning to the end of the ramp, thus treating the whole ramp as a single focusing element. From Eq.(\ref{eqn:bmatrix}) and Eq.(\ref{eqn:comp_mathi}) one can write the normalized transfer matrix for the Twiss function for an injection ramp
\begin{equation}
    R_{inj}
    =
    \begin{bmatrix}
    \\
    s'^2 & -2 ss' & s^2
    \\
    \\
    -c's' & cs'+c's & -cs
    \\
    \\
    c'^2 & -2cc' & c^2
    \\
    \\
    \end{bmatrix}
    \label{eqn:injmatrix}
\end{equation}
and for an extraction ramp
\begin{equation}
    R_{ext}
    =
    \begin{bmatrix}
    \\
    c^2 & -2 cs & s^2
    \\
    \\
    -cc' & cs'+c's & -ss'
    \\
    \\
    c'^2 & -2 c's' & s'^2
    \\
    \\
    \end{bmatrix},
    \label{eqn:extmatrix}
\end{equation}
where $c=c(R)$ and $s=s(R)$.
\begin{figure}[h!]
    \centering
    \centering
    \includegraphics[width=0.5\linewidth]{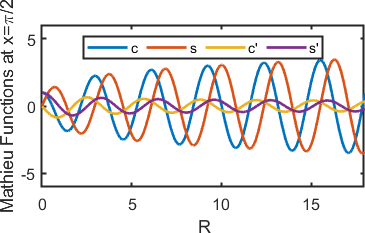}
    \caption{Set of Eq.(\ref{eqn:comp_mathi}) evaluated at $\xi=1$ as a function of $R$.}
    \label{fig:mathi1}
\end{figure}
This kind of matrix is defined for the transport of Twiss parameters that are normalized with respect to $\beta_0$, namely
\begin{equation}
    \begin{bmatrix}
        \beta^*_{1(2)}
        \\
        \alpha_{1(2)}
        \\
        \gamma^*_{1(2)}
    \end{bmatrix}
    =R_{inj(ext)}
        \begin{bmatrix}
        \beta^*_{2(1)}
        \\
        \alpha_{2(1)}
        \\
        \gamma^*_{2(1)}
        \end{bmatrix};
\end{equation}
where $\beta^*=\beta/\beta_0$ and $\gamma^*=\gamma\beta_0$. The subscript index 1 refers to the bunch parameters at the end of the ramp while the index 2 refers to the parameters in the plateau. Since Mathieu functions are transcendental, it is preferable to find an approximated form that allows to handle Eq.(\ref{eqn:injmatrix},\ref{eqn:extmatrix}) in a more simple way. From Fig. \ref{fig:mathi1}, one can notice that the normalized Mathieu transfer elements have a sinusoidal behavior of period $\pi$, together with a slow evolution of the envelope. The envelope of $c,s$ increases with $R$ while the counterpart for $c',s'$ decreases. The function $c$ has a phase delay of approximatively $\pi/2$ with respect to $s$, as well as $c'$ with respect to $s'$. In the same way the Mathieu transfer elements present a phase delay with respect to their derivatives. Further, from Liouville's theorem we have that the transfer matrix is unimodular~\cite{wiedemann2015particle}, namely
\begin{equation}
    cs' - sc' =1.
    \label{eqn:Liouville}
\end{equation}
Taking into account all these properties, one can assume the following approximated form for these equations:
\begin{equation}
    \begin{split}
        c=&aR^b \cos(2R+\omega)\\
        s=&aR^b \sin(2R+\omega)\\
        c'=&-\frac{1}{a(\cos^2\omega-\sin^2\omega)} R^{-b} \sin(2R-\omega)\\
        s'=&\frac{1}{a(\cos^2\omega-\sin^2\omega)} R^{-b} \cos(2R-\omega).\\
    \end{split}
    \label{eqn:mathi2}
\end{equation}
Inserting Eq.(\ref{eqn:mathi2}) into condition Eq.(\ref{eqn:Liouville}) leads to $\cos(2\omega)=\cos^2\omega-\sin^2\omega$, meaning that Liouville's theorem is satisfied by Eq.(\ref{eqn:mathi2}) for any value of $\omega$. In order to evaluate this phase delay, an useful information can be retrieved by plotting the quantity $(c^2+s^2)(c'^2+s'^2)$, shown in Fig.(\ref{fig:mathi3}). As one can notice, this function saturates quickly at $2$ for values of $R>5$, whereby the consistency of the approximation holds if $\omega$ is such that $(c^2+s^2)(c'^2+s'^2)=2$. Since from Eq.(\ref{eqn:mathi2}) follows that $(c^2+s^2)(c'^2+s'^2)=1/\cos^2(2\omega)=2$, $\cos(2\omega)=\sqrt{2}/2$ or $\omega=\pi/8$.
In a similar fashion, it is possible to fit the values of $a$ and $b$ taking into account the following equation
\begin{equation}
   2 a R^b=\sqrt{c^2+s^2}+\sqrt{\frac{2}{{c'}^2+{s'}^2}},
\end{equation}
where we are considering the average between the two functions $\sqrt{c^2+s^2}$ and $\sqrt{2/({c'}^2+{s'}^2)}$ since taking into account only one term of the sum gives lightly different fit results. The fit is shown in Fig.(\ref{fig:mathi4}) and the resulting parameters are $a=1/4$ and $b=\sqrt{3}$.
The final result for the approximation is
\begin{equation}
    \begin{split}
        c(R)\approx&\ \ \ \ \ \sqrt{3}R^{1/4}\cos(2R+\pi/8)\\
        s(R)\approx&\ \ \ \ \ \sqrt{3}R^{1/4}\sin(2R+\pi/8)\\
        c'(R)\approx&-\sqrt{\frac{2}{3}}R^{-1/4}\sin(2R-\pi/8)\\
        s'(R)\approx&\ \ \ \ \ \sqrt{\frac{2}{3}}R^{-1/4}\cos(2R-\pi/8).\\
    \end{split}
    \label{eqn:mathiapp}
\end{equation}
\begin{figure}[h!]
    \centering
    \includegraphics[width=0.5\linewidth]{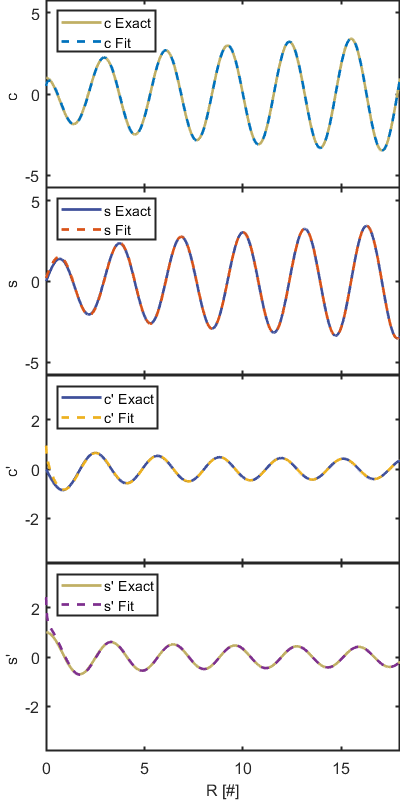}
    \caption{Comparison of the numerical solution for Mathieu functions with the approximation from Eqs.(\ref{eqn:mathiapp}).}
    \label{fig:mathi6}
\end{figure}
\begin{figure}[h!]
    \centering
    \includegraphics[width=0.5\linewidth]{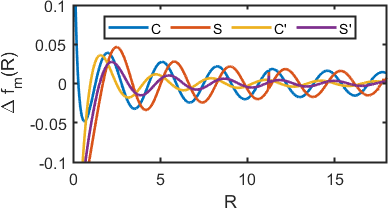}
    \caption{Difference between exact and approximated solutions for Mathieu functions (numerical result minus approximate result).}
    \label{fig:mathi7}
\end{figure}
\begin{figure}[h!]
    \centering
    \includegraphics[width=.5\linewidth]{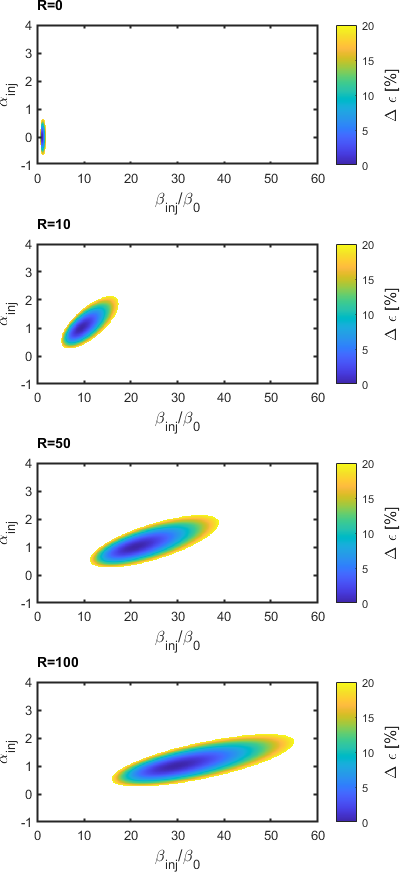}
    \caption{Emittance growth as a function of $\beta_{inj}/\beta_0$ and $\alpha_{inj}$ for different values of $R$. The presence of a ramp greatly relaxes the matching conditions, broadening the stability region where the emittance increase is lower than $20\%$.}
    \label{fig:emit}
\end{figure}
A comparison between Eqs.(\ref{eqn:comp_mathi}) evaluated at $\xi=1$ and Eqs.(\ref{eqn:mathiapp}) as functions of $R$ is shown in Fig.(\ref{fig:mathi6}) and the corresponding difference between the two expressions is shown in Fig.(\ref{fig:mathi7}). The agreement is quite good for $R>1$ and improves for increasing values of $R$.
\section{Stabilizing effect of plasma ramp}
The presence of ramp, as previously stated, helps to relax the matching conditions for the $\beta$-function. If a beam doesn't meet perfect matching conditions, Eq.(\ref{eqn:injmatrix}) can be applied in order to retrieve $\beta^*=\beta(0)/\beta_0$ and $\alpha=\alpha(0)$. If these values are not exactly $1$ and $0$, the envelope will oscillate inside the plateau, causing an emittance growth due to betatron dephasing~\cite{mehrling2012transverse}. The maximum expected emittance growth due to betatron dephasing can be evaluated through the following equation
\begin{equation}
    \varepsilon_{fin}=\frac{\varepsilon_{inj}}{2}\left(\frac{1+\alpha^{2}}{\beta^*}+\beta^*\right)
    \label{eqn:meflo}
\end{equation}
Fig.(\ref{fig:emit}) reports the increase of the stability valley as a function of $R$ if we impose an emittance increase within 20\% of the initial value. What one can deduce from this result is that the presence of the ramp dampens the envelope oscillations that are occurring inside the ramp itself, leading to a condition where the mismatching on the plateau is lower in a much wider range of Twiss parameters. This result is consistent and complementary with the claim performed by Dornmair et al.~\cite{dornmair2015emittance} that the introduction of a plasma ramp dampens the oscillations of the centroid of the beam due to a transverse injection misalignment of a witness bunch respect to the plasma wake and limits the emittance growth.
\section{Comparison with a realistic ramp}
\begin{figure}[t!]
    \centering
    \includegraphics[width=.48\linewidth]{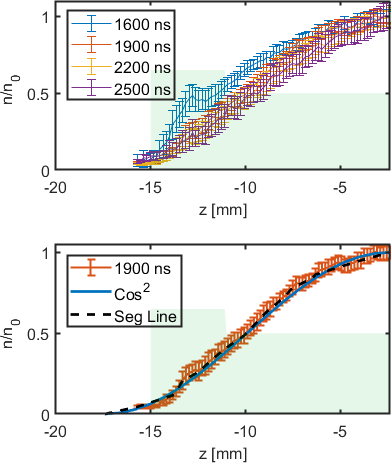}
    \caption{Plasma density measurements in the tapered capillary. Capillary profile is schematized in the background, considering the y-axis as millimiters. On the top it is shown the normalized plasma profile measured at several times after the discharge. After \SI{2}{\micro\second} the profile stabilizes at a squared cosine like plasma profile. On the bottom it is shown the measurement at \SI{1900}{\nano\second} delay, together with a squared cosine fit and an approximation of the line made of 20 segments.}
    \label{fig:dens}
\end{figure}
\begin{figure}[b!]
    \centering
    \includegraphics[width=.5\linewidth]{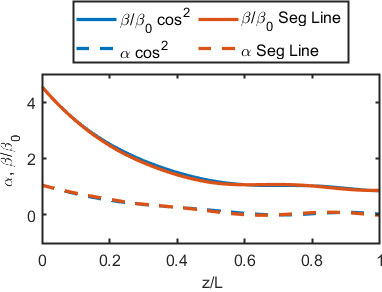}
    \caption{Twiss functions evolution of a \SI{500}{\mega\electronvolt} bunch inside the squared cosine plasma ramp and the segmented plasma ramp shown in Fig.(\ref{fig:dens}). The longitudinal coordinate on the x-axis is normalized respect to the ramp length. The injection conditions of this bunch was evaluated by means of Eq.(\ref{eqn:mathiapp}) and the normalized Twiss parameters at the end of the ramp are $-0.02<\alpha<0.02$ and $\beta/\beta_0\approx0.85$.}
    \label{fig:realramp}
\end{figure}
The assumption of a squared cosine shape was motivated by their smoothness properties that makes them good andidates for representing a realistic experimental setup. The ability to effectively control the plasma density, creating the desired shape, is of primary importance for any possible application of the present work. As previously discussed, ramps naturally arise from discharge capillary generated plasma and they can be manipulated by modifying the capillary geometry. With a proper tapering of the capillary tips, it is already possible to obtain a squared cosine plasma ramp. In Fig. (\ref{fig:dens}) several measurements of the plasma density profile of a capillary with a step tapering are shown. In this particular configuration, the capillary is \SI{3}{\centi\meter} long in total, with the gas entering in a single inlet located at the center. The diameter is \SI{1}{\milli\meter} in the capillary center and becomes \SI{1.3}{\milli\meter} on the edges in a \SI{4}{\milli\meter} length. Plasma density is measured at various times after the discharge occurred and the peak density is in the range \SI{1e17}{\per\cubic\centi\meter} - \SI{6e17}{\per\cubic\centi\meter}. This range of densities is too high for the purposes of EuPRAXIA working point~\cite{assmann2020eupraxia}, but still provides an indication that it is possible to engineer the ramp shape at any density. In Fig.(\ref{fig:realramp}) it is shown the evolution of the normalized Twiss parameters for the ramps shown in Fig.(\ref{fig:dens}). The plateau density was set at \SI{1e16}{\per\cubic\centi\meter} and the Twiss parameters at the injection are evaluated by means of Eqs.(\ref{eqn:mathiapp}). The evolution for both ramps is very consistent, meaning that small deviation from the cosine shape do not introduce major variations in the beam dynamics. The matching is not perfect, since the outcoming normalized Twiss parameters at the plateau are of the order of $\alpha\approx0.02$ and $\beta/\beta_0=0.85$. The expected emittance growth for this kind of bunch, according to Eq.(\ref{eqn:meflo}), is below $1.4\%$; therefore, from all practical purposes, this can be considered an optimal matching.
\section{Conclusions}
In this paper a wide-ranging study regarding the treatment of plasma ramps in linear optics approximation has been presented. In order to uniquely define the focusing strength of a plasma ramp, the ion column model has been adopted together with the assumption of negligible acceleration in the ramp, deriving a linear dependency between the focusing strength and the plasma density as a function of the longitudinal coordinate. The insertion of the ramp shape into the Hill's equation evidenced that the treatment of plasma ramps with an even functional form is more simple since there is a solid demonstration that it is always possible to find for this kind of ramp a solution that satisfy the matching condition at the plateau. The choice of a ramp with squared cosine shape has been performed since this function is continuous, derivable and even. This kind of ramp is non-adiabatic for every value of length, leading to the necessity of a solution for the differential equation without any adiabatic approximation. The resulting differential equation results to be equivalent to a Mathieu differential equation, that has been widely studied in literature, but none of the cases of study could be reconducted to the framework of the paper itself. The solutions given by the Mathieu functions were fully satisfying, showing an high degree of agreement not only with the numerical solution of envelope equation, but with a numerical simulation of a witness bunch injected in the wake of a driver inside plasma. This provided a validation for the use, in the proposed framework, of the ion column model with negligible deceleration for the treatment of plasma ramp. A further treatment of this case of study led to the choice of treating the ramp as a single focusing element, studying the normalized even and odd solutions of the Mathieu differential equation at the end of the ramp only. An approximation of this kind of solutions was found, leading to the definition of the Mathieu transfer matrix for the plasma ramp. This empirical approximation assumed the character of an analytical approximation after the evaluation of the parameters. The fact that the parameters can be expressed as functions of integer numbers and that this approximation results to be progressively more precise with the increase of $R$, suggests that could be possible to derive a solid and analytical demonstration of the Eqs.(\ref{eqn:mathiapp}). Anyhow, this kind of demonstration lies beyond the aim of this work. Beyond the great simplification of the evaluation of the matching conditions in presence of a ramp with the squared cosine shape, a succesful use of the approximated equations has been performed, evaluating in a purely analytical way that one of the effects of the plasma ramp is to dampen the betatron oscillations on the plateau even in mismatched cases. Finally, the experimental feasibility of this kind of ramps have been shown, comparing the transverse evolution of a bunch both in a ramp with a squared cosine shape and a segmented profile that was measured experimentally. The convergence of the results has shown that it is possible to design and realize a plasma ramp that matches the behavior of the squared cosine plasma ramp and that it is possible as well to find an optimal matching with the only use of the proposed matching equations. In conclusion, the theoretical approach that has been performed in this paper shows great potential and a wide range of possible application in the design of plasma based accelerators.
\section{Acknowledgements}
This project has received funding from the European Union´s Horizon Europe research and innovation program under grant agreement No. 101079773.
\vspace{6pt}
\bibliographystyle{iopart-num}
\bibliography{biblio}
\end{document}